\documentstyle[12pt]{article}
\begin{document}
\vspace*{10pt}
\hspace*{15em}\[
\hspace*{15em}\begin{array}{l}
\hspace*{15em}\mbox{MPI-PhT/94-40}\\
\hspace*{15em}\mbox{June~1994}
\hspace*{15em}\end{array}\]
\vspace*{5pt}
\begin{center}
{\sc {\bf ELECTROWEAK SYMMETRY BREAKING WITH NON-UNIVERSAL SCALAR}}\\
{\sc {\bf SOFT TERMS AND LARGE $\tan\beta$ SOLUTIONS}}
\end{center}
\vspace*{5pt}
\begin{center}
M.Olechowski\footnote{$^)$Address: Institute of Theoretical Physics, Warsaw
University.}$^{),}$
\footnote{$^)$Partially supported by the Polish Committee of
Scientific Research.}$^)$
and S.Pokorski\footnote{$^)$On leave from the Institute of Theoretical Physics,
Warsaw University}$^{)}$
\end{center}
\vspace*{10pt}
\begin{center}
{\it Max-Planck Institute for Physics}\\
{\it F\"ohringer Ring 6, 80805 Munich, Germany}
\end{center}
\vspace*{10pt}
\begin{center}
ABSTRACT
\end{center}

We discuss radiative electroweak symmetry breaking with
non-universal
scalar masses at the GUT scale. Large $\tan\beta$ solutions are investigated
in detail and it is shown that qualitatively new (as compared to the
universal case) solutions exist, with much less correlation between
soft terms. We identify two classes of non-universalities which
give solutions with $A_o\simeq B_o\simeq M_{1/2}\simeq O(M_Z)$, $\mu
\simeq O(0.5 m_o)$, $m_o>>M_Z$ and $A_o\simeq B_o\simeq M_{1/2}
\simeq\mu\simeq O(M_Z)$, $m_o\ge M_Z$, respectively. In each case,
after imposing gauge and Yukawa coupling unification, we discuss
the predictions for $m_t$, $m_b$ and the spectrum of supersymmetric
particles. One striking consequence is the possibility
of light charginos and neutralinos which in the
second option can be higgsino-like.
Cosmological constraint on the relic abundance
of the lightest neutralino is also included.

\newpage
\indent The mechanism of radiative electroweak symmetry breaking
in supersymmetric models [1] has been extensively studied in the literature
. The $SU(2)\times U(1)$ gauge symmetry is broken by quantum
corrections to the tree level potential which is postulated at
the GUT scale. The quantum corrections are included by means of the
renormalization group evolution of the parameters of the
lagrangian from the GUT scale to the electroweak scale. Most of the
work [2] has  so far been done under the assumption that the soft
supersymmetry breaking parameters: gaugino masses, scalar masses
and trilinear couplings have at the GUT scale some universal
values $M_{1/2},~m_o$ and $A_o$, respectively. Then, with two
additional parameters: $B_o$ (the soft Higgs mixing term) and
$\mu_o$ (the Higgs mixing in the superpotential),  the model
depends on five free parameters. However, exact universality of
the soft mass terms at the GUT scale is neither
phenomenologically necessary nor theoretically sound. From the
phenomenological point of view, the usual argument based on the
smallness of FCNC does not constrain mass terms which are
diagonal in the family space and even for the off--diagonal
terms the constraint is weaker than usually believed [3]. On the
theoretical side, one may argue that exact universality is
realized at the string scale but no longer at th GUT scale,
where one can envision several types of non--universal
corrections to the soft terms.

Radiative electroweak breaking is expected to be sensitive to the
non--universal corrections to the soft terms in those regions of
the parameter space where the breaking with universal terms
requires a large degree of correlation (fine tuning) among them.
This is the case for the large $\tan\beta$ solutions [4,5] and the
purpose of this letter is to study such solutions in the presence
of non--universal scalar soft mass terms.
Some aspects of non-universal breaking terms have already been
addressed in ref.[6].

The issue of the "large $\tan\beta$" vacuum has ben discussed
for some time [4,7,8] as the possible explanation to the breaking
of the $SU(2)_V$ symmetry of the quark masses:
\[
\tan\beta\equiv\frac{v_2}{v_1}\cong m_t/m_b
\]
with
\[
h_t\simeq h_b
\]
where $v_1,~v_2$ are the vacuum expectation values of the two
Higgs fields which couple to the down and up quarks,
respectively, and $h_t$ and $h_b$ are Yukawa couplings.
Two interesting questions emerge in this context. One is
this: can large breaking of the $SU(2)_V$ by the vacuum be
obtained by amplification (via the mechanism of radiative
breaking) of weak effects such as e.g.  different
hypercharge assignement of the up and down quarks and squarks or
weak $SU(2)_V$ breaking by the soft scalar mass terms at the GUT
scale (or small deviation from $h_t=h_b$ etc.)? Another question
is whether with large $\tan\beta$ one can also explain in a
natural way the gradual restoration of the $SU(2)_V$ symmetry
for the lighter generations, but it will not be addressed here.

Special attention
to the large $\tan\beta$ scenario has been
given in the framework of the models with underlying
$SO(10)$ symmetry [8], which {\it predict}
the equality of the top and bottom Yukawa couplings at the GUT
scale.

In a recent paper [5] a detailed study has been performed of
radiative electroweak breaking with complete gauge and Yukawa unification
and universal soft mass terms at the GUT scale. It has been
found that for large values of $h_t$ (as required for the heavy
top quark) solutions with large $\tan\beta$ are obtained only
for strongly correlated values of the GUT scale parameters.

This can be easily understood in a qualitative way by studying
the tree level scalar potential and the 1--loop RG evolution of
its parameters from the $M_Z$ scale to the GUT scale. For easy
reference we recall the main points here. The Higgs potential
\begin{eqnarray}   
V&=~m^2_1H_1^{\dagger}H_1+m^2_2H^{\dagger}_2H_2&-~m_3^2\left( H^{\dagger}
_1~i\tau_2~H_2+h.c.\right)\nonumber\\
&&+~\mbox{quartic terms}
\end{eqnarray}
has, for large $\tan\beta$ values, two characteristic features.
It follows from the minimization conditions that
\begin{equation}   
\hbox{a})~~~~~~~~~~m^2_2\simeq - \frac{M_Z^2}{2}~~~~~
\end{equation}
and
\begin{equation}   
\hbox{b)}~~~~~~~~~~m^2_3\simeq\frac{M_A^2}{\tan\beta}\simeq 0
\end{equation}
with
\begin{equation}   
M_A^2\simeq m_1^2+m^2_2>0.
\end{equation}
Equations (2) and (4) combined together give a useful constraint
on the low energy parameters
\begin{equation}   
m^2_1-m^2_2>M_Z^2.
\end{equation}
\\
Equations (2) and (3) are the two main constraints on the
parameters of the scalar potential, which are characteristic for
large $\tan\beta$ solutions.

The next information we need is the running of the parameters
from the GUT scale to the $M_Z$ scale. For universal soft
supersymmetry breaking parameters at the GUT scale and for
$ Y_t=Y_b=Y$ ($Y_{t,b}=
\frac{h^2_{t,b}}{4\pi}$)  and large (such that
$Y(M_Z)$ is close to its infrared quasi--fixed point value
$Y_f$, as required by the heavy top quark), the approximate
solutions to the RG equations read[5]:
\begin{equation}   
m_{H_1}^2\simeq m_{H_2}^2=m^2_o+0.5~M^2_{1/2}-\frac{3}{7}\Delta m^2
\end{equation}
where $m_i^2=\mu^2+m^2_{H_i}$,
\begin{eqnarray}   
\Delta m^2&\simeq &3m_o^2\frac{Y}{Y_f}-4.6~A_oM_{1/2}\frac{Y}{Y_f}\left(
1-\frac{Y}{Y_f}\right)\nonumber\\
&+&A_o^2\frac{Y}{Y_f}\left( 1-\frac{Y}{Y_f}\right) +M^2_{1/2}\left[
14\frac{Y}{Y_f}-6\left( \frac{Y}{Y_f}\right)^2\right]
\end{eqnarray}
and
\begin{equation}   
\mu^2=2\mu_o^2\left( 1-\frac{Y}{Y_f}\right)^{6/7}.
\end{equation}
We observe that for large values of Y the masses $m^2_{H_1}$ and
$m^2_{H_2}$ tend to become large and negative for increasing values
of $m_o$ and/or $M_{1/2}$.
For the running of the soft supersymmetry breaking bilinear and
trilinear couplings we have:
\begin{equation}   
A_t\simeq A_o\left( 1-\frac{Y}{Y_f}\right) -M_{1/2}\left( 4.2-2.1\frac{Y}{Y_f}
\right),
\end{equation}
\begin{equation}   
B\simeq \delta(Y) + M_{1/2}\left( 2\frac{Y}{Y_f} - 0.6\right)
\end{equation}
with
\begin{equation}    
\delta(Y) = B_o - \frac{6}{7}~\frac{Y}{Y_f}~A_o.
\end{equation}
Finally, the squark masses read
\begin{eqnarray}    
&&m^2_U\simeq m^2_D=m_o^2 + 6.7~M^2_{1/2} - \frac{2}{7}\Delta m^2,\\
&&m^2_Q\simeq m_o^2 + 7.2~M^2_{1/2} - \frac{2}{7}\Delta m^2.
\end{eqnarray}
It is clear that in the approximation (6) the condition (5),
necessary for the proper symmetry breaking, is not satisfied. However,
eq.\ (6) neglects small differences in the running of the two
Higgs masses which follow from the different hypercharges of the
right top and bottom squarks, from the difference in the running
of the bottom and top Yukawa couplings (equal at the GUT scale) and
from the effects due to the $\tau$ lepton Yukawa. As can be
inferred from the RG equations, the inclusion of those
effects leads to the following result:
\begin{equation}     
m^2_1 - m^2_2 = \alpha M_{1/2}^2 +\beta m_o^2
\end{equation}
where $\beta<0$ due to the $Y_{\tau}$ effects, $\alpha>0$ due to
the other effects mentioned above and both are $O(0.1)$ in
absolute values. With the condition (5) we see now that
large $\tan\beta$
solutions must be driven by large values of $M_{1/2}$:
\begin {equation}     
M_{1/2}>\frac{M_Z}{\sqrt{\alpha}}~~~
,~~~~~M_{1/2}>\sqrt{\frac{|\beta|}{\alpha}} m_o
\end {equation}
Also, it follows from eqs. (2), (6) and (7) that the low energy
superpotential parameter $\mu$ (or its GUT value $\mu_o$; see
eq.\ (8)) is strongly correlated with $M_{1/2}$.
For large $M_{1/2}$, large values of $\mu^2$ are needed to cancel
large negative values of $m^2_{H_2}$ and to satisfy the condition (2).
Indeed, in the limit $Y_t\rightarrow Y_f$ we get
\[      
\mu^2\approx 3M^2_{1/2}.
\]
Finally, in the limit $Y\rightarrow Y_f$, eq.\ (3) and (10)
give (remember $m^2_3\equiv B\mu$):
\begin{equation}    
\delta(Y)\equiv B_o~-~{\frac{6}{7}A_o}\approx~-1.4~M_{1/2}.
\end{equation}
In summary, in the model with universal soft supersymmetry breaking terms
large $\tan\beta$ solutions to radiative
electroweak symmetry breaking  are characterized by large $(\gg
M_Z)$ and linearly correlated values of the parameters
$M_{1/2},~\mu$ and $\delta(Y)$, with the constraint $M_{1/2}>m_o$.
Numerical calculations [5] based on the 2--loop RGE for the gauge
and Yukawa couplings, and one--loop equations for the Higgs and supersymmetric
mass parameters with all the leading supersymmetric threshold
corrections included (bottom--up approach of ref.[9]) confirm the
above qualitative considerations.

Another question one must address in the context of the large
$\tan\beta$ scenario, with $Y_t=Y_b=Y_{\tau}$ at the GUT scale,
is about the supersymmetric threshold corrections to the bottom
mass. We refer the reader to ref.[10,5] for a detailed discussion
and here repeat only the main points. Fig.1 (taken from ref.[5])
shows the predictions for the top quark (pole) mass as a function of the
strong coupling constant $\alpha_s(M_Z)$ under the assumption of
the gauge  and Yukawa coupling unification
for several values of the bottom mass, when
supersymmetric corrections to $m_b$ are ignored. The shaded
region shows the $(m_t,~\alpha_s)$ values consistent with those
assumptions and with  $m_b$  in the experimental range
$(4.9\pm 0.3)$ GeV. Let us suppose experiment will confirm the
values of $m_t$ and $\alpha_s$ in that range. Then the question
emerges: is the parameter space obtained from radiative breaking
consistent with the assumed small supersymmetric corrections to the $b$ mass
? Another option is also open: experimental data for $m_t$ and
$\alpha_s$ will place us outside the shaded region of Fig.1,
i.e. in the region where the corresponding bottom mass (with
$Y_t=Y_b=Y_{\tau}$ and {\it no} SUSY corrections) is larger than
the experimental value $(4.9\pm 0.3)$ GeV. Then large SUSY correction
to $m_b$ will be necessary to reconcile the measured values of
$m_b,~m_t,~m_{\tau}$ and $\alpha_s$ with  full unification of the gauge and
Yukawa couplings. One of the main results of ref. [5] is that in
the model with universal soft terms the supersymmetric loop
correction to the bottom mass (mainly gluino and higgsino exchange):
\begin{equation}    
{\delta m_b\over m_b}\simeq\left( {\alpha_3\over 3\pi}~{M_{\tilde{g}}
\mu\over m^2_{\tilde q}}~+~{Y_t\over 8\pi}~{A_t\mu\over m^2_{\tilde q}}\right)
\tan\beta
\end{equation}
are large, 0(20--50\%). This follows from the discussed above
pattern of radiative electroweak breaking and selects the second
option as the only one consistent with this mechanism.
Another important physical feature of those solutions is heavy
superpartner spectrum, with only the Higgs pseudoscalar remaining
light [5].

Strong correlations between parameters signal a high degree of fine tuning
which is needed for large $\tan\beta$ solution with {\it
universal} soft terms. Of course, it is not excluded that such,
very regular correlations may follow from the future theory of
the soft supersymmetry breaking terms. However, at the present
purely phenomenological level they are triggered by the
smallness and the signs of the coefficients $\alpha$ and $\beta$
in eq.\ (14), which have their origin
in the effects mentioned after eq.\ (13).
It is important to realize that solutions
consistent with eq.\ (14) and (5) are very unstable with respect
to small perturbations, if they can reverse the sign of $\beta$:
for $\beta >0$, qualitatively new solutions become possible,
with $M_{1/2}\simeq 0$ and
\begin{equation}     
m_o >~{M_Z\over\sqrt{\beta}}
\end{equation}
{}From inspection of the RGE, it is easy to see that there are (at
least) three ways to reverse the sign of $\beta$ by perturbing
slightly the universal boundary conditions at the GUT scale:
\begin{eqnarray*}     
&&\hbox{a)}~~~~~~~~~~Y_t>Y_b\\
\hbox{and/or}~~~~~&&\hbox{b)}~~~~~~~~~~
(m^U_o)^2 \equiv (1+\delta_U)m^2_o>(1+\delta_D)m^2_o \equiv (m^D_o)^2\\
\hbox{and/or}~~~~~&&\hbox{c)}~~~~~~~~~~
(m^{H_1}_o)^2 \equiv (1+\delta_1)m^2_o>(1+\delta_2)m^2_o \equiv (m^{H_2}_o)^2
\end{eqnarray*}
Already O(20-30\%) perturbation is sufficient to reverse the sign of
$\beta$.
The possibility of small deviation from the equality $Y_t=Y_b$ has already
been discussed in ref. [5]. Here we focus on the non-universal
scalar masses.

In the presence of non-universal boundary values for the scalar
masses the coefficients of the $m^2_o$ terms in eqs.(6) and (7)
are appropriately modified. For a qualitative insight one can
use the following approximate formula
(in the limit $Y\rightarrow Y_f$):

\begin{equation}   
\Delta\beta =
\left[{2\over 5}+{3\over5}\left(1-{Y\over Y_f}\right)^{5/7}\right]
\left(\delta_1-\delta_2\right)
+{3\over 5}\left[1-\left(1-{Y\over Y_f}\right)^{5/7}\right]
\left(\delta_U-\delta_D\right).
\end{equation}
It takes values O(0.1) for $(m^{H_1}_o)^2-(m^{H_2}_o)^2$ or
$(m^U_o)^2-(m^D_o)^2$ of order O(20\%)$m_o^2$ and O(0.5) for O($m^2_o$),
respectively.

It should be stressed that the pattern (18) of
symmetry breaking is now determined by the positive sign of the
coefficient $\beta$ and essentially does not depend on the
origin of this sign. However, as we shall see, there are important
properties of the solutions which do depend on the character of the
non-universality
at the GUT scale.

With $\beta>0$, it follows from eq.\ (18) that there exists a
new class of solutions
with $M_{1/2}\simeq 0$ and $m_o\gg M_{1/2}$.
The allowed region in the parameter space $(M_{1/2},~m_o)$
is shown  in Fig.2.
In this and the other figures we present the results of our
numerical calculations (based on the method developed in ref.[9]
and ref.[5]) for
a) $m_t=170$ GeV, $\tan\beta=49$ and with the GUT
scale values $m^2_U=1.2 m_o^2$, $m^2_D=0.8 m_o^2$;
b) $m_t=180$ GeV, $\tan\beta=53$, $m^2_{H_2}=0.7 m_o^2$;
c) $m_t=180$ GeV, $\tan\beta=53$, $m^2_{H_1}= 2.0m_o^2$, $m^2_{H_2}=1.5 m_o^2$;
where $m_o$ is the universal mass of the other scalars.
Here we refer to the overall
regions in Fig.2  where the solutions to radiative breaking exist: the only
constraints which are taken into account are the present experimental limits on
the sparticle masses and the requirement that the lightest neutralino
is the lightest supersymmetric particle. The other regions in Fig.2
will be discussed shortly.

Further properties of the solutions and certain classification
of non-universalities follows from the equation for $\mu^2$
modified by the non-universal
scalar terms. In the limit $Y\rightarrow Y_f$ it reads:
\begin{equation}  
\mu^2 = -\frac{M_Z^2}{2} + \Delta_m m^2_o + \Delta_M M^2_{1/2}
+\frac{3}{7}\frac{Y}{Y_f}\left(1-\frac{Y}{Y_f}\right)A_o\left(A_o-4.6M_{1/2}\right)
\end{equation}
where
\[
\Delta_M \simeq 6\frac{Y}{Y_f}-\frac{18}{7}\left(\frac{Y}{Y_f}\right)^2-0.5
{}~~,~~~~~
\Delta_m \simeq \frac{9}{7}\frac{Y}{Y_f} - 1 + \Delta_s
\]
with
\[
\Delta_s = - \delta_2
+\frac{3Y}{14Y_f}\left(\delta_U+\delta_D+\delta_2+\delta_1\right)
+\frac{3}{10}\left[1-\left(1-\frac{Y}{Y_f}\right)^{5/7}\right]
 \left(\delta_U-\delta_D+\delta_2-\delta_1\right)
\]
Several observations follow from eq.\ (20): The strong
($M_{1/2},~\mu$) correlation triggered by eq.\ (20) in the limit
of universal scalar terms
disappears, due to the importance of both $m_o$ and $M_{1/2}$
contributions, depending on the particular solution.
Nevertheless, the values of $\mu^2$ remain correlated
with $m_o$ or $M_{1/2}$ in the limit
$m_0 \gg M_{1/2}$ and $M_{1/2} \gg m_o$,
respectively.
In addition, in the limit $m_0 \gg M_{1/2}$
 which is of interest for us
the values of $\mu^2$
depend,  contrary to the sign of $\beta$,
on the pattern of the deviation from universality in the scalar masses.
In the limit $Y\simeq Y_f$ we obtain the following classification:
\\
A) $\mu^2 >
\Delta_M M^2_{1/2}$ ~~for~~
$\Delta_s > -2/7$
\\
B) $\mu^2<
\Delta_M M^2_{1/2}$ ~~for~~ $\Delta_s<-2/7$
\\
It is clear that with $\beta>0$, as in eq.\ (19), the first
option can be realized eg. for $(m_o^{H_2})^2<m_o^2$ or $(m_o^U)^2>m_o^2$,
with the masses of the other scalars at $m_o^2$. Generically the values
of $\mu$ remain then large $\mu \gg M_Z$. This case is illustrated by the
examples (a) and (b) of our numerical calculations.

The possibility (B) (with $\beta>0$) can be realized when $m_o^D<m_o^U<m_o$
or/and $m_o<m_o^{H_2}<m_o^{H_1}$, for large enough deviations from
universality.
The parameter $\mu$ can then be
arbitrarily small
\footnote{With the same mechanism one can obtain $\mu\simeq 0$ also
when $\beta<0$, i.e. when radiative breaking is driven by large $M_{1/2}$.
However, one needs then $\delta_2\gg 1$ and, in addition, large $M_{1/2}$
brings us back to the strong $(\delta(Y),M_{1/2})$ correlation present
in the universal case.}.
Thus, radiative breaking can be driven by $m_o \geq M_{1/2}\simeq O(M_Z)$,
with $\mu\simeq O(M_Z)$ and uncorrelated with $m_o$. In our numerical
calculations example (c) illustrates this case. We stress that the above
approximate considerations are only meant as a qualitative quideline
to the complete numerical calculations.

We can now understand all the details of the full solution regions
in Fig.2 as well
as the solution regions in the planes $(\mu,M_{1/2})$ and $(\mu,m_o)$
shown in Fig.3 and 4, respectively. In case (A) the $(m_o,M_{1/2})$
region is bounded from below by the constraints (15) and (18).
The values of $(\alpha , \beta)$ are generically
$(O(0.2),O(0.02))$ and $(O(0.2),O(0.01))$ for
case (a) and (b), respectively. For
large values of $M_{1/2}$ we have a lower bound on $m_o$ from the
requirement $m_{\tilde l}>m_{\chi_1^o}$ (the lightest neutralino).
The lower bound on $M_{1/2}$ is the experimental one obtained from
the CDF limit on the gluino mass.
In case (B) additional constraints follow from the condition
$\mu^2>(50 \hbox{GeV})^2$ (experimental limit) which for a given $m_o$
gives a lower bound on $M_{1/2}$. The absolute lower bound on $M_{1/2}$ is
therefore higher
than the experimental bound of $\sim$50 GeV.
The plots in Fig.3 and Fig.4 are also nicely consistent
with our qualitative discussion.  The lower limit on $M_{1/2}$
as a function of $\mu$, in case (c), follows from eq.\ (20). The
upper bound for $M_{1/2}$ as a function of $\mu$ in case (a) and (b)
corresponds to the linear correlation present in the universal case.
Comparing cases (a) and (b), we see the effects of approaching
the limit $Y=Y_f$.

The correlation between $\delta(Y)$ and
$M_{1/2}$ remains as given by eq.\ (16). However, for the solutions with
$M_{1/2}\approx O(M_Z)$ eq.\ (16) can be
satisfied with
\begin{equation}       
B_o\sim A_o\sim M_{1/2}\sim O(M_Z)
\end{equation}

We conclude that with non-universal scalar masses we have
the interesting new range (21) of the parameter space
which gives large $\tan\beta$ solutions with
\\
$\mu\ge O(0.5m_o)$, $m_o\gg M_Z$ in case (A)
\\
$\mu\sim O(M_Z)$, $m_o\ge M_Z$ in case (B).

The main source of the potential fine--tuning are the constraints $m^2_2\cong
-{1\over 2}M^2_Z$ and $m^2_3\cong 0$ which have to be satisfied
for varying values of the parameters at the GUT scale. The
degree of fine--tuning can be measured by the derivatives
\begin{equation}     
{m^2_o\over m^2_2}{\partial m^2_2\over \partial m_o^2}\cong {\mu_o^2\over
m^2_2}{\partial m^2_2\over \partial\mu^2_o}\cong O\left( {m^2_o\over
M^2_Z}\right)
\end{equation}

\begin{equation}  
{M^2_{1/2}\over m^2_2}{\partial m^2_2\over\partial
M^2_{1/2}}\cong O\left( {M^2_{1/2}\over M^2_Z}\right)
\end{equation}

\begin{equation}  
{Y_t\over m^2_2}{\partial m^2_2\over\partial Y_t}\cong O\left( {m^2_o\over
M^2_Z}\right)
\end{equation}

\begin{equation}   
{M_{1/2}\over\tan\beta}{\partial\tan\beta\over\partial M_{1/2}}\cong
{\delta(Y)\over\tan\beta}{\partial\tan\beta\over\partial\delta(Y)}\cong
{\delta(Y)\over B}\cong {M_{1/2}\over B}
\end{equation}
\begin{equation}   
{Y_t\over\tan\beta}{\partial\tan\beta\over\partial Y_t}\cong
{A_o\over B} \cong {\mu A_o\over m^2_3}
\end{equation}
where  eqs.\ (6--11) and (15) have been used in the calculation.
(We have also conservatively assumed that the derivative of $Y_t(M_Z)$ over its
values at $M_{GUT}$ is 1 and neglected in eqs.\ (22--26) several
small terms.) The derivatives in eqs.\ (22),(23), and (24) are of the
order $\ge O(\frac{1}{\alpha})$ or $\ge O(\frac{1}{\beta})$ and reflect
some fine tuning needed to satisfy eq.\ (2) for solutions with universal
and non-universal soft terms, respectively. The derivative in eq.\ (25)
(and similarly for (26)) is $O(M_{1/2}\tan\beta /\beta m_o)$ and
O($\tan\beta$ /$\alpha$) for non-universal and universal case, respectively.
In the latter case its value is O(500) whereas in the former -- depending
on the values of $\beta$ and $m_o$ -- it can be even as low as O(10).
At this point our conclusion is that, with non--universal scalar
mass terms, radiative electroweak breaking has qualitatively new
solutions for large $\tan\beta$, with moderate fine tuning.

There are, however, two additional important constraints which have
to be considered:
we have not addressed yet the question of consistency
of radiative breaking with full unification of gauge and Yukawa
couplings and, secondly, the relic abundance of the stable neutralino
should not overclose the Universe.
As we know from ref.[5], the first  requirement is a strong
constraint for the model with universal soft terms due to the
potentially large supersymmetric corrections to the bottom mass.
We recall again that for ($m_t,~\alpha_s$) values in the shaded
region of Fig.1 the $\delta m_b/m_b$ correction given in eq.\
(17) has to be small whereas for $(m_t, ~\alpha_s$) values
outside that region the correction has to be non--negligible and
in the right range, in order to be consistent with coupling
unification. For  given values of $(m_t,~\alpha_s)$ (or
equivalently $m_t$ and $\tan\beta$; see Fig.\ 1) it is,
therefore, necessary to study the parameter space obtained after
radiative breaking {\it and} constrained by the requirement of
the right magnitude for the correction $\delta m_b$. In
particular, it is interesting to understand if there are now
consistent solutions for the shaded region in Fig.1 i.e. with
$\delta m_b/m_b\leq 10\%$, say. For a qualitative discussion, we
can consider the first term in eq.\ (17) (this is a conservative
estimate as the second term partially cancells the first one;
see eq.\ (9)).
The condition

\begin{equation}    
{\delta m_b\over m_b}\approx 0.5 {M_{\tilde{g}}\mu\over m^2_{\tilde{q}}}
< 1/10
\end{equation}
can be fulfilled in case (A) (which gives $\mu\sim O(m_{\tilde q})$)
for $m_{\tilde q}>5 M_{\tilde g}$. Taking into account the experimental
limit $M_{\tilde g}>$ 150 GeV we need $m_{\tilde q}>O(1-2)$ TeV and
correspondingly $m_o\sim O(2-4)$ TeV (since $M_{\tilde g}\ll m_{\tilde q}$,
the squark masses must be almost entirely determined by $m_o$).
The regions constrained by the condition (27) are marked
in Fig.2, 3 and 4 and are cherecterized by upper bounds
on $M_{1/2}$ for given $m_o$ (i.e.\ $m_{\tilde q}$) or given $\mu$.
Again, comparing (a) and (b) we see the effect of approaching the limit
$Y\rightarrow Y_f$. The regions in ($m_o$, $\mu$) plane
reflect the correlation following from eq.\ (20) in the limit $m_o\gg M_{1/2}$.

In case (B), due to the lower limit on $M_{1/2}$
as a function of $m_o$ shown in Fig.2 (which follows from the
condition $\mu^2>(50 \hbox{GeV})^2$), the squark masses
$m^2_{\tilde q}\sim O(0.2)m^2_o + 4M_{1/2}^2$ are determined by the gluino
contribution and $M_{\tilde g}/m_{\tilde q} \sim O(1)$.
The condition (27) is satisfied for $m_{\tilde q}> O(5\mu)$
which is achieved for $m_{\tilde q}> O(300 \hbox{GeV})$, as
$\mu$ can take small values.
Clearly, eq.\ (27) gives a lower bound on $M_{1/2}$ as a function
of $\mu$ which is seen in Fig.3. But at the same time $\mu$
is given by eq.\ (20) so this bound must be consistent with eq.\ (20).
We obtain then a lower bound on $m_o$ as a function of $M_{1/2}$,
whereas the upper bound is given by the condition $\mu > (50 \hbox{GeV})^2$.
With $\Delta_m<0$, the allowed region is never empty.
This explains the pattern seen in Fig.2. Similarly can be
understood Fig.4. One should stress the importance of the
non--zero gaugino mass in satisfying simultaneously all the
constraints in case (B). In particular it assures the positivity
of the squark masses squared and of the $\mu^2$ for a given deviation
from universality in the scalar masses

It is obvious from Figs.\ 2, 3 and 4 that the constraint (27), by imposing
certain cuts on the otherwise large parameter space with large $\tan\beta$
solutions, requires somewhat more fine tuning. In option (A) it follows
mainly from large values of $m_o$ (eq.\ (22) and (24)) and in case (B)
it is the $M_{1/2}>\hbox{O}(M_Z)$ which is important for eq.\ (25) and (26).
Generically, after the cut (27), the derivatives (22)--(26) are O(100).

The superpartner mass spectra obtained with non--universal scalar masses show
several new and interesting features as compared to the universal case.
Some of them are presented in Figs.\ 5 and 6. The most important aspects
are (we discuss the spectra with eq.\ (27) taken into account):
\begin{itemize}
\item[1.]
{Light charginos and neutralinos; they reflect very small
values of $M_{1/2}$ (case (A)) or $\mu$ (case (B)).
This latter case is particularly interesting as it revives our
interest in light higgsino--like neutralinos.}
\item[2.]
{Clear differences between sparticle spectra in case (A) and (B):
they reflect the available parameter regions in both cases.
We get, respectively
\\
a) very heavy, $\geq$ O(1 TeV), or light, $\geq$ O(100 GeV),
sleptons.
\\
b) very heavy, $\geq$ O(1 TeV), or moderately light, $\geq$ O(300 GeV),
squarks.
\\
c) very light, $\geq$ O(150 GeV), or somewhat heavier, $\geq$ O(300 GeV),
gluinos.}
\item[3.]
{The Higgs sector is characterized by generically light
pseudoscalar $A$. This follows from the equation
$M_A^2=\alpha M_{1/2}^2 + \beta m_o^2 - M_Z^2$,
with small coefficients $\alpha$ and $\beta$.
In the allowed parameter range, both in cases (A) and (B),
the mass $M_A$ can reach the present experimental limit
of about 50 GeV and increases with increasing $m_o$ and/or $M_{1/2}$.
The mass of the light scalar $h$ is strongly correlated with $M_A$
and changes from $M_h\sim 60$ GeV for $M_A\sim 50$ GeV
to $M_h\sim(120-140)$ GeV for $M_A>100$ GeV.}
\end{itemize}

Finally, we discuss the condition $\Omega h^2<1$ for the stable neutralino.
In case (A) it is strongly gaugino--like ($\mu>M_{1/2}$)
and annihilation proceeds mainly through the Higgs pseudoscalar exchange
(all sfermions are heavy). Explicit calculation gives
$M_A\leq$ O(200) GeV for $\Omega h^2<1^{[11]}$.
Therefore we get the constraint
\begin{equation}   
m_o\leq\frac{M_Z\sqrt{6}}{\sqrt{\beta}}
\end{equation}
which has to be satisfied together with the lower bound $m_o\geq$ O(2 TeV).
This constraints $\beta$ to small values $\beta\sim$ O(0.01).
The non--universal boundary conditions chosen for the numerical
calculations give $\beta$ in this range.

In case (B) the stable neutralino is often dominantly higgsino--like
or at least a mixture of gaugino and higgsino. Its annihilation is more
effective and uncorrelated with $M_A$. The nautralino relic abundance
is generically small, $\Omega h^2\sim$ O(0.01--0.1), and the requirement
$\Omega h^2<1$ does not constrain the parameter space.

The discussed solutions to radiative breaking are consistent with
some of the approximate symmetries suggested in ref.\ [10] in order
to stabilize large $\tan\beta$ solutions and to protect $m_b$
from large supersymmetric corrections. In case (A) these are
\begin{equation}   
\frac{A}{m_{\tilde q}} \sim \frac{M_{\tilde g}}{m_{\tilde q}}
\sim \frac{B}{m_{\tilde q}} \sim 0
\end{equation}
and in case (B)
\begin{equation}   
\frac{B}{m_{\tilde q}} \sim \frac{\mu}{m_{\tilde q}} \sim 0.
\end{equation}
However, it is not possible to satisfy simultaneously all the
symmetries of ref.[10]. For instance, in case (B), the result
$B/m_{\tilde q} \approx 0$ is obtained by some moderate
cancellation between $A_0\sim B_0\sim M_{1/2} \sim$ O($M_Z)$,
with $M_{\tilde g}/m_{\tilde q} \sim A/m_{\tilde q} \sim$ O(1).
It is due to the fact that, for large $\tan\beta$, $B \ll M_Z$
and only solutions with $M_{1/2} \approx 0$
would be free of this type of fine tuning.
On the other hand, in the first case $\mu/m_{\tilde q}\sim$ O(1) and
there is moderate fine tuning in $m_2^2$ between $\mu^2_o$ and $m^2_o$.

With non--universal scalar terms top quark mass up to the IR
quasi fixed point values, $m_t \sim$ (190--200) GeV, can be
accommodated in the model. In case (A) this can be achieved at the
expense of heavy squarks but in case (B) is as easy as for
$m_t \sim$ (170-180) GeV

In summary, non--universal scalar terms at the GUT scale open
qualitatively new possibilities for radiative elecktroweak
breaking with natural large $\tan\beta$ solutions.
Two different patterns of non--universality can be identified,
with distinct predictions for the particle spectra.
Solutions with $A_o \sim B_o \sim M_{1/2} \sim \mu \sim \hbox{O}(M_Z)$,
$m_o \geq M_Z$ look particularly interesting.
Top quark masses up to the IR fixed point values
(190--200) GeV can be easily accommodated in this scheme.

\vskip1cm
The main results of this paper have been presented in a talk
by SP at the SUSY'94 workshop in Ann Arbor, May 14--17. Related results
were presented at SUSY'94 by H.P. Nilles [12] and A. Pomarol [6].
We would like to thank these people for discussions.

\newpage
\begin{center}
REFERENCES
\end{center}

\begin{enumerate}{\leftmargin0pt}
\item 
{L.E. Ibanez and G.G. Ross, Phys.\ Lett.\ B110 (1982) 215;\\
K. Inoune et al., Prog.\ Theor.\ Phys.\ 68 (1982) 927;\\
L. Alvarez--Gamum\'e, J. Polchinksy and M. Wise,
Nucl.\ Phys.\ B221 (1983) 495;\\
J. Ellis, J. Hagelin, D. Nanopoulos and K. Tamvakis,
Phys.\ Lett.\ B125 (1983) 275;\\
L.E. Ib\'anez and G. L\'opez, Nucl.\ Phys.\ B233 (1984) 411.}
\item 
{For the most recent results see:\\
R. Arnowitt and P. Nath, Phys.\ Rev.\ Lett.\ 69 (1992) 1014;
Phys.\ Lett.\ B287 (1992) 89 and B289 (1992) 368;\\
M. Olechowski and S. Pokorski, Nucl.\ Phys.\ B404 (1993) 590;\\
V. Barger, M.S. Barger and P. Ohmann,
University of Wisconsin--Madison preprint, MAD/PH/801 (1994);\\
G.L. Kane, C. Kolda, L. Roszkowski and J.D. Wells,
Michigan report UM--TH--94--03 (1994).}
\item 
{A. Brignole, L. Ibanez and C. Munoz, preprint FTUAM--26/93\\
D. Choudhury, F. Eberlein, A. K\"onig, J. Louis and S. Pokorski,
in preparation}
\item 
{M. Olechowski and S. Pokorski, Phys.\ Lett.\ B214 (1988) 393;\\
G.F. Giudice and G. Ridolfi, Z.\ Phys.\ C41 (1988) 447;\\
H.P. Nilles, "Beyond the Standard Model", in Proceedings of the 1990
Theoretical Advanced Study Institute in Elementary Particle Physics, p. 633;
Eds. M. Cvetic and P. Langacker, World Scientific;\\
P.H. Chankowski, Diploma Thesis (1990), University of Warsaw;\\
A. Seidl, Diploma Thesis (1990), Technical University, Munich;\\
W. Majerotto and B. M\"{o}sslacher, Z. Phys.  C48, 273 (1990);\\
S. Bertolini, F. Borzumati, A. Masiero and G. Ridolfi, Nucl.\ Phys.\
 B353, 591 (1991);\\
M. Drees and M.M. Nojiri, Nucl.\ Phys.\ B369, 54 (1992);\\
B. Ananthanarayan, G. Lazarides and Q. Shafi, Bartol Research Institute
preprint BA--92--29.}
\item 
{M. Carena, M. Olechowski, S. Pokorski and C. Wagner,
preprint CERN--TH.7163/94 (1994) to appear in Nucl.\ Phys.\ B.}
\item 
{N. Polonsky and A. Pomarol,
University of Pennsylvania preprint UPR--0616--T (1994);\\
R. Hempfling, DESY preprint (4-078 (1994);\\
A. Leyda and C. Munoz, Phys.\ Lett.\ B317 (1993) 82;\\
T. Kobayashi, D. Suematsu and Y. Yamagishi,
Kanazawa report KANAZAWA--94--06, hep--ph 9403330;\\
R. Rattazzi, U. Sardir and L.J. Hall, talk presented at the second
IFT Workshop on Yukawa couplings and the origin of mass, February 1994
Gainesville, Florida, SU--ITP--94/15, hep--ph 9405313;\\
Y. Kawamura, H. Murayama and M. Yamaguchi, DPSU--9402 and LBL--35731,
hep--ph 9406245.\\
M. Carena and C. Wagner, private communication.}
\item 
{P. Krawczyk and S. Pokorski, Phys.\ Rev.\ Lett.\ 60 (1988) 182.}
\item 
{B. Anantharayan, G. Lazarides and Q. Shafi, Phys.\ Rev.\ D44 (1991) 1613;\\
S. Dimopoulos, L.J. Hall and S. Raby,
Phys.\ Rev.\ Lett.\ 68 (1992) 1984; Phys.\ Rev.\ D45 (1992) 4192;\\
G.W. Anderson, S. Raby, S. Dimopoulos and L.J. Hall,
Phys.\ Rev.\ D47 (1993) 3702;\\
G.W. Anderson, S. Raby, S. Dimopoulos, L.J. Hall and G.D. Starkman,
Phys.\ Rev.\ D49 (1994) 3660.}
\item 
{M. Olechowski and S. Pokorski in ref.\ [1].}
\item 
{L.J. Hall, R. Rattazzi and U. Sarid, LBL preprint LBL--33997 (1993).}
\item 
{P. Gondolo, M. Olechowski and S. Pokorski, to appear.}
\item 
{D. Matalliotakis and H.P. Nilles, Munich preprint MPI--PhT/94-39 (1994).}
\end{enumerate}

\newpage
\begin{center}
FIGURE CAPTIONS
\end{center}

\begin{itemize}
\item[Fig. 1.]
Top quark mass predictions as a function of the
strong gauge coupling, within the framework  of
exact unification
of the three Yukawa couplings of the third generation.
The solid lines represent  constant values of the pole
bottom mass $m_b$ (without supersymmetric corrections),
equal to A) 4.6 GeV, B) 4.9 GeV,
C) 5.2 GeV, D) 5.5 GeV and E) 5.8 GeV. The dashed lines
represent constant values of  $\tan\beta$,
equal to a) 40, b) 45, c) 50, d) 55 and e) 60.
The region to the right of the dotted line is
consistent with the unification of  gauge couplings and the
experimental correlation between $m_t$ and $\sin^2\theta_W(M_Z)$.
The long-dashed line represents the infrared
quasi-fixed-point values for
the top quark mass, for which $Y_t(0) \simeq 1$. \\
\item[Fig. 2.]
The regions in ($M_{1/2}$, $m_o$) parameter space which give
radiative electroweak symmetry breaking and consistent with all
experimental limits on sparticle masses.
\\
Short dashed curves:  $m_t=170$ GeV,
$\tan\beta = 49$,
$m^2_Q(0)=1.2m^2_o$,
$m^2_D(0)=0.8m^2_o$.
\\
Long dashed curves:  $m_t=180$ GeV,
$\tan\beta = 53$,
$m^2_{H_2}(0)=0.7m^2_o$.
\\
Solid curves: $m_t=180$ GeV,
$\tan\beta = 53$,
$m^2_{H_1}(0)=2.0m^2_o$,
$m^2_{H_2}(0)=1.5m^2_o$.
\\
The marked areas correspond to $\delta m_b/m_b<10\%$, in each case.
\item[Fig. 3.]
 The same as Fig.\ 2 for ($M_{1/2}$, $\mu$).
\item[Fig. 4.]
 The same as Fig.\ 2 for ($m_o$, $\mu$).
\item[Fig. 5.] The same as Fig.\ 2 for the predictions for chargino and slepton
masses.
\item[Fig. 6.]
 The same as Fig.\ 2 for the predictions for stop and gluino masses.
\end{itemize}

\end{document}